\documentstyle[prl,aps, 12pt]{revtex}

\title{A variational study of the random-field XY model}
\author{T.Garel, ~G.Iori ~and ~H.Orland \\
Service de Physique Th\'eorique\\
CE-Saclay, 91191 Gif-sur-Yvette Cedex\\
France}
\date{\today}

\def\be{\begin{equation}}
\def\th0{\theta_0}
\def\phr{\phi_{\vec r}}
\def\bea{\begin{array}}
\def\ee{\end{equation}}
\def\eea{\end{array}}
\newcommand{\unmezzo} {\frac{1}{2}}
\newcommand{\thetar} {\theta_{\vec r}}
\newcommand{\thetarp} {\theta_{\vec r'}}
\def\vr{\vec r}
\def\vk{\vec k}

\newcommand{\vrp}{\vec r'}
\newcommand{\gti}{\tilde g}
\newcommand{\ddk}{\frac{d^d k}{(2 \pi)^d}}

\newcommand{\guno}{\gamma_1}
\begin{document}
\maketitle
\vskip 1cm
\begin{abstract}
A disorder-dependent Gaussian variational approach is
applied to the $d$-dimensional ferromagnetic XY model in a random field.
The randomness yields a non extensive contribution to the
variational free energy,
implying a random mass term in correlation functions.
The Imry-Ma low temperature result, concerning the existence ($d>4$)
or absence ($d < 4$) of long-range order is obtained in a transparent
way. The physical picture which emerges below $d=4$ is that of a marginally
stable mixture of domains.
We also calculate within this variational scheme, disorder
dependent correlation functions, as well as the probability
distribution of the Imry-Ma domain size.
\end{abstract}
\vskip 3cm\noindent\mbox{Submitted for publication to:} \hfill
\mbox{Saclay, SPhT/95-124}\\ \noindent \mbox{``Physical Review B''}\\
\vskip 1cm\noindent \mbox{PACS: 75.10N, 64.70P, 71.55J} \newpage

\newpage

\section{Introduction}
\label{sec: intro}
The effect of quenched disorder on systems with
continuous symmetry has recently
attracted a lot of theoretical as well as experimental
interest. Experimental realizations include, among others, arrays of
flux-lines in  type II  disordered superconductors,
crystalline surfaces with a disordered
substrate, spin or charge-density waves subject
to random pinning, etc...
Most theoretical studies of this model have focused on the
random-field vortex-free
XY case, which is equivalent to the random phase
sine-Gordon model.
\cite{Cardy_Ostlund,Gold_Houg_Sch,Vill_Fer,Toner_Vic,MPFisher,Korshunov,Giam_LeDou,Tsvelik,Balents_Kardar,Hwa_DFisher,Batrouni_Hwa,Cule_Shapir1,Cule_Shapir2,LeDou_Giam,Nattermann,Toner,Tsai_Shapir}

Recently, a disorder dependent variational approach has been
proposed for this problem \cite{Orland_Shapir}. In this approach, the
disorder enters only through a unique variable $u=\int d \vec x
\cos(2\pi d(\vec x))$, where $d(\vec x)$ is a random phase.
This random variable has a Gaussian
distribution, and for $u<0$, one recovers the results of the replica
Gaussian variational principle, with Parisi symmetry breaking scheme
\cite{Giam_LeDou}.

One major advantage of this
approach is that it is genuinely variational, thus
providing a true upper bound to the
free energy of the system, unlike replica based methods, which are
plagued by the $n=0$ limit.

In this paper, we shall use this variational approach in the framework
of the (full) XY model in a random magnetic field, for which few
results are available \cite{Imry_Ma,Gingras_Huse}. Its Hamiltonian
reads:
\begin{equation}
\label{ham}
{\cal H} = - J \sum_{<\vec r, \vec r'>} \cos (\theta_{\vec r} - \theta_{\vec
r'}) - \sum_{\vec r} h_{\vec r} \cos (\theta_{\vec r} - \phi_{\vec r})
\end{equation}
where $\theta_{\vec r}$ denotes the phase angle of the XY spin $\vec
S_{\vec r}$, $J$ is the ferromagnetic nearest-neighbor coupling
constant, and $(h_{\vec r},\phi_{\vec r})$ are the polar coordinates
of the random field $\vec h_{\vec r}$. The probability distribution of
the (site uncorrelated) random field is Gaussian, given by:
\begin{equation}
\label{proba}
P({\vec h_{\vec r}}) = \left({1 \over 2\pi h^2}\right)^N
\exp\left(-\sum_{\vec r}  {{\vec h_r}^2 \over {2 h^2}}\right)
\end{equation}
where $N= L^d$ is the total number of sites on a $d$-dimensional
hypercubic lattice of linear size $L$ and lattice spacing $a$;
$h$ is the variance of the
random field.

The physics of the Hamiltonian (\ref{ham}) is believed to be well
captured by the Imry-Ma argument, which we briefly summarize \cite{Imry_Ma}.
Consider,
at low temperature,
a magnetized domain of size $L$, with magnetization in the $\theta_0$
direction. To study the stability of such a region, we imagine that we
construct inside a subdomain of size $\xi$, in which the magnetization
is aligned with the average local magnetic field. The
energetic balance reads:
\be
\label{IM}
\Delta E \sim J \xi^{d-2} - h \xi^{d/2}
\ee
where the first term represents the spin-wave distortion energy,
whereas the second represents the magnetic energy gain due to the
random field. Following \cite{Imry_Ma}, we conclude that the lower
critical dimension for the system is $d_c=4$. Above $d_c$, the
energetic cost is prohibitively high, so that such domains of size
$\xi$ cannot exist: ferromagnetic long range order is
stable. On the contrary, for $d < d_c$, the whole system
will break into subdomains of smaller size, so as to maximize its
magnetic energy gain. Consequently, no long range ferromagnetic order
may exist below $d_c$.

Usually, in quenched disordered systems, extensive thermodynamical quantities,
such as the free energy, are identified to their average over
the disorder \cite{Brout}. This can be understood in the
following way: one divides the macroscopic system into mesoscopic
subsystems
, each subsystem corresponding to a particular
disorder configuration. For short-range forces, the free energy is
additive, and thus the total free energy is the sum of the free
energies of the subsystems.
This procedure clearly neglects all correlations or
domain-wall energies between neighboring subsystems. In the present
case, these non-extensive contributions are precisely of the same
order of magnitude than the terms of equation (\ref{IM}). Therefore, in the
following, we will not perform quenched averages, but rather keep the
disorder variables throughout the calculations.

The layout of this paper is the following. In section II, we define the
variational Hamiltonian, and calculate the corresponding variational
free energy. The variational method yields two
solutions.
In this approximation, we find that the transition
temperature is the same for both solutions,
equal to that of the pure system.
In section III,
we discuss the issue of long-range order as a function
of space dimension. We find
long-range ferromagnetic order at low temperature in dimensions
$d>4$. For lower dimensions, the discussion is postponed to the next section.
In section IV,
we discuss the non-extensive corrections to the free energy, and study
the stability of the variational solutions.
For dimensions $2 < d < 4$,  we argue that the physical solution is the
marginally
stable one, shedding light on the Imry-Ma domain picture.
In the conclusion,
we calculate the
probability distribution for the Imry-Ma length, and discuss the issue
of correlation functions.
\section{The variational free energy }
\label{sec: free}
We consider Hamiltonian (\ref{ham}) and its associated variational Gaussian
companion:
\be
\label{h0}
\beta {\cal H}_0 = \unmezzo \sum_{\vr,\vrp} (\thetar - \th0)~ g(\vr -
\vrp)^{-1}(\thetarp - \th0)
\ee
where we have restricted the variational kernel $g$ to be
translationally invariant,
the direction of magnetization $\th0$ is space independent
and $\beta = 1/T$ is the inverse temperature.

The true free energy
$F$ satisfies the bound:
\begin{equation}
\label{bound}
F \le \Phi (\{ \vec h_{\vec r}\}) = F_0 + <{\cal H}-{\cal H}_0>_0
\end{equation}
Using equations (\ref{ham},\ref{h0}), we obtain the disorder
dependent variational free energy:
\be
\label{Phi}
\Phi(\{\vec h_{\vec r}\}) = - \frac{1}{2 \beta} \sum_{\vec k} ~ \ln \tilde
g(\vec k) -
\frac{J}{2} L^d e^{-g(0)} \sum_{\alpha=1}^{2d} e^{g(e_{\alpha})} -
e^{-{g(0)}/2} \sum_{\vec r} h_{\vec r}\cos (\phr -\th0)
\ee
where $\tilde g(\vec k)$ is the Fourier transform of $g(\vr)$,
$\{e_\alpha\}, \alpha=1,...,2d$ denote the lattice unit vectors and
$\{h_{\vr}, \phr \}$ are the polar coordinates of the random field
$\vec h_{\vr} $ at site $\vr$.

The variational equation with respect to $\th0$ reads:
\be
\label{th0}
e^{-{g(0)}/2} \sum_{\vec r} h_{\vec r}\sin (\phr -\th0) = 0
\ee
If we define $\phi$ as the polar angle of the total magnetic field $\vec H =
\sum_{\vec r} \vec h_{\vec r}$, we see that the solutions to
(\ref{th0}) are
$\th0 =
\phi$, corresponding to a magnetization aligned with the total
magnetic field $\vec H$, and $\th0= \phi+\pi$ corresponding to a magnetization
opposite to $\vec H$.

The minimization of $\Phi(\{\vec h_{\vec r}\})$
with respect to $\gti (\vk)$ then
yields, for the two solutions:
\be
\label{var}
{1 \over \beta \gti (\vk)} = J \sum_{\alpha=1}^{2d} (1 -
e^{ik_\alpha}) e^{-(g(0)-g(e_\alpha))}\pm e^{-{1\over 2} g(0)}
L^{-d} \sum_{\vec r} h_{\vec r}\cos (\phr -\phi)
\ee
the $+$ (resp. $-$) sign corresponding to $\theta_0 = \phi$ (resp.
$\theta_0 = \phi + \pi$). For the sake of simplicity, we shall stick
below to a unique notation $\gti (\vk)$ for the two solutions. In the
following, we
shall refer to these solutions as the ($+$) solution (for
$\theta_0 = \phi$) and the ($-$) solution (for $\theta_0 = \phi + \pi$).

Defining the components $h_x =\sum_{\vec r} h_{\vr} \cos(\phr)\ ,\ h_y
=\sum_{\vec r} h_{\vr} \sin(\phr)$ , we see that:
\be
\label{H}
\sum_{\vec r} h_{\vec r}\cos (\phr -\phi) = H
\ee
The total magnetic field $\vec H$ is a random variable, and the
distribution of its modulus is given by the central limit theorem.
Defining the positive random variable $u$ by:
\be
\label{u}
 H = L^{d/2}\ h u
\ee
we have:
\be
\label{CLT}
P(u) =  u \exp (-u^2/2)
\ee
Each disorder configuration
is thus specified by a single positive random
variable $u$. The $u$-dependent free energy per site (the upper sign
corresponding to the ($+$) solution, and the lower sign to
the ($-$) solution).
\be
\label{free}
\phi(u) = {\Phi(u) \over L^d} = - \frac{1}{2 \beta} \int \ddk ~ \ln \tilde
g(\vec k) -
{J \over 2} e^{-g(0)} \sum_{\alpha=1}^{2d} e^{g(e_{\alpha})} \mp
e^{-{g(0)}/2} L^{-d/2}h u
\ee
where we have replaced $\sum_{\vk}$ by $L^d \int \ddk$ (see below).

Using the symmetry of the problem, we set $ \gamma_1 = g(0) -
g(e_{\alpha})$ independent of $\alpha$. Summing over $\alpha$, we get:
\be
\label{guno}
d \guno = \sum_{\alpha = 1}^{2d} (g(0) - g(e_{\alpha}))=
\sum_{\alpha=1}^{2d} \int \ddk \gti (\vk) (1 - e^{ik_\alpha})
\ee
Inserting equation (\ref{var}) in equation (\ref{guno}), we obtain the
(disorder dependent) critical temperature through:
\be
\label{crit}
2 d J e^{- \gamma_1} \gamma_1 = T \mp  g(0) e^{- \unmezzo g(0)}
L^{-d/2}h u
\ee
For large system size ($L \to \infty$), we see that the second term of
the r.h.s is always negligeable, and is therefore identical for the
two solutions. The critical temperature is the same
as in the pure system $T_c^{\rm pure} = {2 d J \over e}$, where
$e=2.718..$. The quantity $\guno$ varies from $0$ (at $T=0$) to $1$
(at $T_c^{\rm pure}$).

As is clear from equation (\ref{var}),
the form of $\gti(\vk)$ and therefore of $\phi(u)$ is very different
for the two solutions. For the ($+$) case,
the replacement of $\sum_{\vk}$ by $L^d
\int \ddk$ is straightforward, whereas the ($-$) case
requires a more careful treatment, due to the existence
of poles in $\gti(\vk)$.

\section{Long range order and the Imry-Ma argument}
\label{sec: IM}
\subsection{The ($+$) solution}
\label{sec: IMp}
In this case, equation (\ref{var}) shows that:
\be
\label{gr}
 \beta g (\vr) = \int \ddk {e^{i \vk \vr}\over{J \over 2} \sum_{\alpha=1}^{2d}
(1 -
e^{ik_\alpha}) e^{-(g(0)-g(e_\alpha))}+ e^{-{1\over 2} g(0)}
L^{-d/2} h u}
\ee
implying (for $d>2$) the existence of long range order both in
the angular and spin variables.
Indeed, we have:
\be
\label{fluctuations}
<(\theta_{\vec r} - \theta_0)^2 > = g(0)
\ee
\be
<S_{\vec r}^x> = \cos \phi \ \exp(-{g(0) \over 2})
\ee
and $g(0)$ is finite (infrared convergent) for $d>2$.

For $d=2$, one finds,
as in the pure case, algebraic order, but with different exponents.
For example, the total magnetization reads:
\be
\label{mag}
M^x = \frac{1}{N} \sum_{\vr} <S_{\vr}^x>= \cos \phi \ e^{-{g(0)\over 2}} \sim
\left(\frac{a^2}{L\xi}\right)^{\frac{\guno/\pi}{2-\guno/\pi}} \cos \phi
\ee
with $\xi = {J \over hu}e^{-\guno}$,
the pure system behaving as
\be
M_{pure}^x \sim
\left(\frac{a}{L}\right)^{\guno/\pi}
\ee
Note the appearance of the ($u$ dependent) Imry Ma length scale $\xi$ in
equation
\ref{mag}. We do not investigate further the case ($+$), since it
will be shown, in section \ref{sec: pic}, that the
physically relevant case ($2 < d < 4$) is the ($-$) one.
\subsection{The ($-$) solution}
\label{sec: IMm}
In this case, one has to be more careful in taking the continuous
limit in (\ref{free}), since the poles may yield a finite contribution
to thermodynamic quantities (in close analogy to the Bose
condensation). For instance, we may write (for small wave vectors):

\be
\label{smallk}
\beta J L^d g(0) = \sum_{\vk} \frac{e^{\guno}}{\vk^2 - \mu^2}
\ee
where
\be
\label{mu2}
\mu^2 = u \frac{he^{\guno-g(0)/2}}{JL^{d/2}}
\ee
In equation
(\ref{smallk}), there may arise a singular contribution from the $2d$
smallest wavevectors
 ${\vk = \frac{2\pi}{L}(1,0,0,...)}$ (plus permutations). One therefore has
\be
\label{poles}
\beta J L^d g(0) = \frac{2de^{\guno}}{\frac{4\pi^2}{L^2} - \mu^2} +
L^d \int_{k>\frac{2\pi}{L}} \ddk \frac{e^{\guno}}{\vk^2 - \mu^2}
\ee
A detailed  discussion of equation (\ref{poles}) requires separate
treatments for $d >4$, for $2< d <4$ and for $d=2,\ 4$.
\subsubsection{Dimension $d > 4$}
\label{sec: d4}
According to (\ref{mu2}), we have $\mu^2 << L^{-2}$. We may thus
neglect the $\mu^2$ contribution in the denominators of (\ref{poles}),
yielding a finite magnetization:
\be
M^x = \frac{1}{N} \sum_{\vr} <S_{\vr}^x> =- e^{-\frac{AT}{J(d-2)}} \cos \phi
\ee
where $A = {1 \over 2^d \pi ^{d/2}} { e^{\guno} \over \Gamma (d/2) a^{d-2}}$
\subsubsection{Dimension $ 2 <d < 4$}
\label{sec: d24}
Since the integral in (\ref{poles}) is infrared (IR) convergent, it may
be neglected compared to the pole contribution. In this case, one
finds:
\be
\beta J L^d g(0) \simeq \frac{2de^{\guno}}{\frac{4\pi^2}{L^2} - \mu^2}
\ee
This last equation shows that (when $L \to \infty$), $\mu$ is very
close to $2\pi /L$. Using equation (\ref{mu2}), we get:
\be
g(0) = (4-d) \ln \frac{L}{\xi_d(u)}
\ee
where
\be
\label{xid}
\xi_d(u) = \left( \frac{4\pi^2 J e^{-\guno}}{h u}\right)^{2/(4-d)}
\ee
Since $g(0)$ diverges for large $L$, we conclude that there is no long
range order in this case.

\subsubsection{Dimensions $d = 2$ and $d=4$}
\label{sec: d2}
For $d=2$,
one may easily see that both terms in the r.h.s. of
equation (\ref{poles}) are of the same order of magnitude. Indeed, the
integral has a logarithmic divergence, and
performing
the calculation, we obtain the same result as above, namely:
\be
\label{g0}
g(0) = 2 \ln \left( { L \over \xi_2(u)} \right)
\ee
with
\be
\label{xi2}
\xi_2(u) = \left( \frac{4\pi^2 J e^{-\guno}}{h u}\right)
\ee
For $d=4$, one gets:
\be
g(0) = 2 \ln \left({h u \over {4 \pi^2 J e^{-\guno}}} \right)
\ee
implying disorder induced fluctuations in the magnetization.

\section{Marginal stability  and finite size corrections}
\label{sec: pic}
Whenever there are several solutions to the variational equations, one
ought to pick the one with the lowest free energy.
In our case, it is
easily seen that the extensive part is the same for both solutions,
and that the difference shows up only in non-extensive corrections.
\subsection{Dimension $2 < d <4$}
To leading order, the variational free energy reads,
\be
\label{ffree}
\Phi_{+}(u) = F_0 L^d - A_{+}\ h u\ L^{d/2}
\ee
\be
\Phi_{-}(u) = F_0 L^d + (4-d)\ A_{-}\ J e^{-\guno} \ L^{d-2}\ \ln \left({L
\over
\xi_d(u)} \right)
\ee
where
\be
\label{cons}
F_0 =  -{ 1 \over 2 \beta} \ln \left ( {e^{\gamma_1} \over \beta J} \right )
-d J e^{- \guno} + {1 \over 2 \beta} \int \ddk \ln (\vk ^2)
\ee
and $A_{+} $ and $A_{+} $ are positive constants, independent of $L$, depending
on geometrical factors as well as on the temperature of the system.

The above selection criterion
holds only if the two minima are not degenerate. In the opposite case,
the selection rule is provided for by the fluctuations
since they may
contribute to the extensive part of the free energy.
In other words,
one has to check for the stability of the solutions, with respect to
variations of $\gti (\vk)$ and local variations $\delta_{\vr}$
of $\theta_0$.

We consider first the $\delta_{\vr}$  fluctuation
contribution to the free energy and get:
\be
\label{delta}
\Delta F = \sum_{\vr} \left( {J\over 4} e^{-\gamma_1} \sum_{\alpha=1}^{2d}
(\delta_{\vr} - \delta_{\vr + \vec {e_\alpha}})^2 +{1\over 2}\ e^{-g(0)/2}
\ h_{\vr}\ \cos(\phi_{\vr} - \theta_0)\ \delta_{\vr} ^2 \right)
\ee
The positivity of $\Delta F$ is determined by the spectrum of the
kernel in (\ref{delta}). This kernel is analogous to that of a
tight-binding model in a random potential:
\be
\label{schrod}
\left(-{J\over 2} \Delta_{\vr \vr'} + V_{\vr} \right) \Psi = \lambda \Psi
\ee
where
\be
\label{pot}
V_{\vr} = e^{-g(0)/2} e^{\gamma_1} h_{\vr} \cos(\phi_{\vr} - \theta_0)
\ee
It is easy to see that ${V_{\vr}}$ is a random Gaussian distributed
potential with:
\be
\label{mean}
\overline {V_{\vr}} = e^{-g(0)/2} \sqrt{\pi \over 2} h L^{-d/2} \to 0
\ee
\be
\overline{V_{\vr} V_{\vrp}} = e^{-g(0)} h^2 \delta (\vr- \vrp)
\ee
where the bar stands for an average over the random field.

Using standard perturbation theory to second order, the ground state
energy of (\ref{schrod}) can be expanded around the zero energy mode
$\vk =0$ as:
\be
\label{gs}
E_0 = <\vec 0 | V |\vec 0> - \sum_{\vk \ne \vec 0} { |<\vk | V |\vec
0>|^2 \over J {\vk}^2/2 }
\ee
where $|\vk>$ denote the normalized plane waves.

Since $V$ is a random potential of mean given by (\ref{mean}), the first term
of the r.h.s. of (\ref{gs}) is of order $\overline{V} \sim L^{-d/2}$.
The second order term is negative and finite, and thus the energy $E_0$ is
negative, leading to a fluctuation induced instability of the
finite magnetization $(+)$ solution.

This instability is in agreement with the Imry-Ma picture, which shows
that for $d <4$, a uniformly magnetized system is unstable with
respect to domain formation.

On the contrary, for the zero-magnetization $(-)$ solution, the
disordered potential vanishes. The spectrum is that of a free
particle, implying the existence of zero energy modes.

Furthermore, one may check that the $(-)$ solution is stable with respect to
variations of $\gti (\vk)$, since its stability properties with
respect to these variations are the same as the pure system.

Putting all these results together, we may conclude that
 the $(-)$ solution is marginally stable, and thus appears
as the physical solution.

\subsection {Dimension $d > 4$}
It is easily seen that both solutions having finite magnetization,
(aligned or opposite to the field), they are unstable. This is again
in accord with the Imry-Ma argument, which shows that for $d>4$, the
magnetization (i.e. $\theta_0$) is not determined by the random field, but
rather by a
standard infinitesimal uniform field. This in turn ensures the
stability of the uniform magnetisation with respect to small
fluctuations.

\subsection {Dimensions $d=2$ and $d=4$}
According to section \ref{IM}, for $d=2$, both solutions are marginally stable.
This is a borderline case, due to the existence of a Kosterlitz-Thouless
transition
in the pure system.

For $d=4$, it can be seen that the random potential of equation
(\ref{schrod}) is not Gaussian. This dimension is likely to be a
special dimension for the localization problem. This fact is also
present in the framework of the $d \to d-2 $ dimensional reduction
theory \cite{Imry_Ma_Aha,Parisi_Sourlas}.

\section{Conclusion}
We have presented a disorder dependent variational method for the
full $XY$ model in a random field. We recover the Imry-Ma results
concerning
the existence ($d>4$) or absence ($2<d<4$) of long range order. In
particular, we find in the latter case, a
 variational solution which is marginally stable with respect to local
magnetization rearrangements. Using eqs.(\ref{CLT}) and (\ref{xid}),
this solution can be further
characterized by the probability distribution of the Imry-Ma domain
length $\xi_d$, which reads
\be
\label{pxi}
P(\xi_d)={(4-d) \over 2} {B  \over \xi_d^{5-d}} \exp(-{B \over 2 \xi_d^{4-d}})
\ee
 with $B=({4 \pi^2 J e^{-\guno} \over h})^2$. Eq.(\ref{pxi}) shows in
particular, that the Imry-Ma domains are characterized by multiple
length scales; for d=3, the average domain size is found to be
divergent, whereas the most probable domain size is of order $B$.

In a similar way, one may deduce the spin-spin correlation function.
We have
\be
K(\vec r)=<\vec S_0 \vec S_{\vec r}>= e^{-(g(0)-g(\vec r))}
\ee
Considering only the case $2<d<4$, we have
\be
K(\vec r)= e^{-\left(({4-d \over d}) \ w({x_{\alpha} \over L})\ln{L \over
\xi_d(u)} \right)}
\ee
where
\be
 w({x_{\alpha} \over L})=\sum_{\alpha=1}^d (1-\cos{2 \pi x_{\alpha} \over L})
\ee
and $x_{\alpha}$ denote the coordinates of $\vec r$. The disorder
dependence of the correlation function is contained in the Imry-Ma
length scale $\xi_d(u)$:
 one may therefore get its probability
distribution. Here we just point out two limiting cases for the
disorder dependent spin-spin correlation function. At small distances
($x_{\alpha}<<L$), $K(\vec r)$ is Gaussian. At large distances ($x_{\alpha}
\sim L$), $K(\vec r)$ decreases like a power law
\be
K(\vec r) \sim ({\xi_d(u) \over r})^{w(4-d) \over d}
\ee
where $w$ is the value of $w({x_{\alpha} \over L})$ for $x_{\alpha}
\sim L$. This behaviour is in broad agreement
 with the results of (i) a real space renormalization group
\cite{Vill_Fer}  (ii) a
replica variational calculation \cite{Giam_LeDou}, both calculations pertaining
to a
vortex free model. The full XY model in a random field has
been recently studied \cite{Gingras_Huse} by Monte Carlo calculations for
$d=2,3$,
indicating (for $d=3$) the possibility of a phase transition to a
pinned vortex free phase. Our variational approach emphasizes the
existence of a probability distribution for the Imry-Ma domain size,
which renders the comparison with this work rather delicate. Finally,
our results should be of interest in other Imry-Ma like situations
\cite{Aiz_Weh,Ber_Hui}.
\newpage


\begin{references}
\bibitem[]{}\begin{center}
{\bf REFERENCES}
\end{center}
\bibitem{Cardy_Ostlund}
J.L. Cardy and S. Ostlund, Phys. Rev. B {\bf 25}, 6899 (1982).
\bibitem{Gold_Houg_Sch}
Y.Y. Goldschmidt and A. Houghton, Nucl. Phys. B {\bf 210}, 175 (1982);
Y.Y. Goldschmidt and B. Schaub, {\it ibid}. B {\bf 251}, 77 (1985).
\bibitem{Vill_Fer}
J. Villain and J. F. Fernandez, Z. Phys. B {\bf 54}, 139 (1984).
\bibitem{Toner_Vic}
J. Toner and D. P. DiVincenzo, Phys. Rev. B {\bf 41}, 632 (1990).
\bibitem{MPFisher}
M. P. A. Fisher, Phys. Rev. Lett. {\bf 62}, 1415 (1989).
\bibitem{Korshunov}
S. E. Korshunov, Phys. Rev. B {\bf 48}, 3969 (1993).
\bibitem{Giam_LeDou}
T. Giamarchi and P. Le Doussal, Phys. Rev. Lett. {\bf 71}, 1530 (1994).
\bibitem{Tsvelik}
A.M. Tsvelik, Phys. Rev. Lett. {\bf 68}, 3889 (1992).
\bibitem{Balents_Kardar}
L. Balents and M. Kardar, Nucl. Phys. B {\bf 393}, 480 (1993).
\bibitem{Hwa_DFisher}
T. Hwa and D.S. Fisher, Phys. Rev. Lett. {\bf 72}, 2466 (1994).
\bibitem{Batrouni_Hwa}
G. G. Batrouni and T. Hwa, Phys. Rev. Lett. {\bf 72}, 4133 (1994).
\bibitem{Cule_Shapir1}
 D. Cule and Y. Shapir, Phys. Rev. Lett. (in press).
\bibitem{Cule_Shapir2}
D. Cule and Y. Shapir, Phys. Rev. B (in press).
\bibitem{LeDou_Giam}
P. Le Doussal and T. Giamarchi, Phys. Rev. Lett. {\bf 74}, 606 (1995).
\bibitem{Nattermann}
T. Nattermann, Phys. Rev. Lett. {\bf 64} 2454 (1990).
\bibitem{Toner}
J. Toner, Phys. Rev. Lett. {\bf 67}, 2537 (1991);
{\it ibid.} {\bf 68}, 3367 (1990).
\bibitem{Tsai_Shapir}
Y.-C. Tsai and Y. Shapir, Phys. Rev. Lett. {\bf 69}, 1773 (1992) ;
 Phys. Rev. E (in press).
\bibitem{Orland_Shapir}
H. Orland and Y.Shapir, Europhys. Lett. {\bf 30}, 203 (1995).
\bibitem{Imry_Ma}
Y.Imry and S.-k. Ma, Phys.Rev.Lett. {\bf 35}, 1399 (1975).
\bibitem{Gingras_Huse}
M.J.P.Gingras and D.A.Huse, ``Topological defects in the random field
XY model and randomly pinned vortex lattices'', preprint.
\bibitem{Brout}
R.Brout, Phys.Rev. {\bf 115}, 824 (1959).
\bibitem{Imry_Ma_Aha}
Y.Imry, S.-k. Ma and A. Aharony, Phys.Rev.Lett. {\bf 37}, 1364 (1976).
\bibitem{Parisi_Sourlas}
G.Parisi and N.Sourlas, Phys.Rev.Lett. {\bf 43}, 744   (1979).
\bibitem{Aiz_Weh}
M.Aizenman and J.Wehr, Phys.Rev.Lett.  {\bf 62}, 2503 (1989)
\bibitem{Ber_Hui}
K.Hui and A.N.Berker, Phys.Rev.Lett.  {\bf 62}, 2507 (1989)
\end{references}
\end{document}